\documentstyle[12pt]{article}
\begin{document}
\title{{\bf SPACE FOR BOTH \\NO-BOUNDARY AND TUNNELING\\
	QUANTUM STATES \\OF THE UNIVERSE}
\thanks{Alberta-Thy-03-97, gr-qc/9704017}}
\author{
Don N. Page
%\thanks{CIAR  Cosmology Fellow}
\thanks{Internet address:
don@phys.ualberta.ca}
\\
CIAR Cosmology Program, Institute for Theoretical Physics\\
Department of Physics, University of Alberta\\
Edmonton, Alberta, Canada T6G 2J1
}
\date{(1997 April 7)}

\maketitle
\large
\begin{abstract}
\baselineskip 14.5 pt

	At the minisuperspace level of homogeneous models,
the bare probability for a classical universe
has a huge peak at small universes for the Hartle-Hawking
`no-boundary' wavefunction, in contrast to the suppression
at small universes for the `tunneling' wavefunction.
If the probability distribution is cut off at the Planck
density (say), this suggests that the former quantum state is
inconsistent with our observations.
For inhomogeneous models in which stochastic inflation can occur,
it is known that the idea of including a volume factor
in the observational probability distribution
can lead to arbitrarily large universes' being likely.
Here this idea is shown to be sufficient
to save the Hartle-Hawking proposal
even at the minisuperspace level
(for suitable inflaton potentials), by giving it
enough space to be consistent with observations.

\end{abstract}
\normalsize
%\baselineskip 15 pt

%\section*{I.  INTRODUCTION AND MAIN RESULT}

	Various remarkable features of our universe,
not implied by the dynamical laws that seem to govern its evolution,
strongly suggest that its state is not random but highly special.
Examples of these special features are
the large size and low curvature of the present universe,
the approximate homogeneity and isotropy of the matter distribution
on the largest scales that we can see today,
and especially the high degree of order
in the early universe that has enabled entropy to increase,
as described by the second law of thermodynamics.

	Two leading proposals for
special quantum states of the universe
are the Hartle-Hawking `no-boundary' proposal 
\cite{HVat,HH,H,HalHaw,P,Hal,HLL}
and the `tunneling' proposal of Vilenkin, Linde, and others
\cite{V1,V2,TL,TZS,TR,TV}.
In toy models incorporating
presumed approximations for these proposals,
they both seem to lead to low-entropy early universes
and so might explain the second law of thermodynamics.
If a suitable inflaton is present in the effective low-energy
dynamical theory, and if sufficient inflation occurs,
both proposals seem to lead to
approximate homogeneity and isotropy today.
However, it has been controversial whether both proposals
do indeed predict sufficient inflation for the large
present size of the universe.

	In particular, in the minisuperspace approximation
of using only Robertson-Walker geometries
and a single homogeneous inflaton scalar field,
the tree-level or zero-loop probability densities
for the two proposals have the opposite signs in the exponent
of the Euclidean action $S_{\rm E}$
(itself inversely proportional to the inflaton potential $V(\phi_0)$
at the nucleation value $\phi_0$ of the inflaton field $\phi$,
when the nucleation is via a Euclidean four-dimensional hemisphere):
 \begin{equation}
 P_{\rm NB} = e^{-2S_{\rm E}} = e^{\pi a_0^2} = \exp{[{3 \over 8V(\phi_0)}]}
 \label{eq:1}
 \end{equation}
for the Hartle-Hawking no-boundary proposal, and
 \begin{equation}
 P_{\rm T} = e^{-2|S_{\rm E}|} = e^{+2S_{\rm E}} = e^{-\pi a_0^2}
 = \exp{[-{3 \over 8V(\phi_0)}]}
 \label{eq:2}
 \end{equation}
for the tunneling proposal, where
 \begin{equation}
 a_0 = a_0(\phi_0) = [8\pi V(\phi_0)/3]^{-1/2},
 \label{eq:3}
 \end{equation}
is the radius of the Euclidean four-dimensional hemisphere
that is a solution of the Einstein equations with an
energy density $V(\phi_0)$ that is the same in all frames
(i.e., a stress-energy tensor $T_{\mu\nu} = - V g_{\mu\nu}$),
and where I am using Planck units throughout ($\hbar = c = G = 1$).
(If one thus distinguishes the tunneling proposal from
the no-boundary proposal by the sign of the exponent
in the corresponding probability, then by this particular criterion
one should exclude \cite{V1,V2}
as descriptions of the tunneling proposal, since they
gave the same sign of the exponent as the no-boundary proposal.
There can also be differences of opinion 
as to which of the remaining
references refer to the tunneling proposal, 
since they actually make slightly different proposals.  
For this paper I shall refer to them all 
as the tunneling proposal, 
even though I am really referring to 
a class of different proposals.)

	During inflation the inflaton potential $V(\phi)$ decreases
to some particular value where inflation ends,
so if one presumes that one has a realization of
the universe configuration in which the probability density
is roughly maximized, then at this level the tunneling proposal
seems to favor the maximum amount of inflation possible,
whereas the no-boundary proposal seems to favor the minimum amount.
Typically the maximum amount of inflation is infinite
(e.g., when the inflaton field in unbounded,
or when the inflaton potential has a maximum),
so in this regard the tunneling proposal seems consistent
with observations, which are themselves apparently consistent
with an arbitrarily large universe.
However, the minimum amount of inflation,
just sufficient for it to be called inflation,
is very small \cite{GR},
leading to a universe that would recollapse
long before it got large enough to be consistent
with our observations of the universe.

	One might conclude that the no-boundary proposal
has thus been refuted by observations.
However, before rejecting it by such a simple-minded argument,
one should look for possible correction factors.
Therefore, I shall describe various corrections
that have been proposed (though not in historical order).

	One correction is the inclusion of higher-loop effects.
At the one-loop level, it has been shown \cite{Bar}
that there might be either an enhancement
or a damping of the probability density
at very large inflaton potentials at the initial nucleation
of the universe as a quasi-classical entity.
However, this seems grossly insufficient to overcome
the enormous exponentially large probability density
of the no-boundary proposal at unacceptably low values
of the potential (leading to far too little inflation),
at least if one restricts attention to one-loop effects
where they are not expected to be dominated by higher-loop effects
(i.e., at densities below the Planck value).

	Furthermore, if the one-loop effects lead to a damping
of the probability density, this can make the tunneling proposal
give the maximum probability density at an initial value
of the potential that is finite
and leads to a finite amount of inflation.
If this finite amount is too small to fit the observations
(which seem to require at least about 60 or so e-folds of inflation
\cite{Lin90,KT}),
then the tunneling proposal, as well as the no-boundary one,
would seem to be in trouble.
However, it is easy for the unknown parameters
entering the inflaton potential and the one-loop correction
to be such that even the finite amount of inflation
thus predicted for the tunneling proposal would be sufficient
(typically much more than sufficient) to fit with observations
\cite{Bar}.

	Thus one-loop effects, if they do not make the problem worse
for both proposals, seem to leave the no-boundary proposal
at the same disadvantage it apparently has relative
to the tunneling proposal from the zero-loop calculation.

	A second correction is to consider the total probability
rather than just the probability density.
In other words, instead of simply asking where
the peak of the probability density is, one should, for each proposal,
compare the total probability for sufficient inflation
with the total probability for insufficient inflation.
Here is the first consideration by which the no-boundary proposal
might be as viable as the tunneling proposal \cite{HP}.
For although the unnormalized probability density
of the no-boundary proposal has an utterly enormous peak
at tiny amounts of inflation, one can easily see that
if the nucleating value of the inflaton field is unbounded above,
and if loop or other effects do not damp the probability density
at these large values (where the zero-loop approximation has
the probability density tending to a constant,
assuming that the potential either diverges
or tends to a constant in this limit),
then the integral of the probability density
over this infinite range of the value of the inflaton field
gives a diverging unnormalized probability for sufficient inflation.
This swamps the exponentially large but finite unnormalized probability
for insufficient inflation, giving a prediction
that the no-boundary proposal leads to sufficient inflation
(actually an arbitrarily large amount of inflation)
with unit normalized probability.

	The usual objection to this argument for saving
the no-boundary proposal is that for it to work
for an inflaton potential that rises indefinitely
(and not unnaturally slowly) for arbitrarily large values
of the inflaton field, one must assume that the probability density
is not unduly damped for large values of the potential
that greatly exceed the Planck-density,
which is roughly the largest density
where the zero-loop approximation
might be expected to be rather reliable.
It is sometimes said that one should put a cutoff
on the probability distribution at the Planck density,
in which case the integral over the probability density
at lower values is grossly insufficient to overbalance
the huge peak at minimal values for inflation
in the no-boundary proposal.

	The first part of this objection is indeed valid,
that the total-probability solution to the apparent difficulty
of the no-boundary proposal does seem to require suitable physics
in the Planck regime, which we certainly do not yet understand.
Thus this solution is only a possible solution,
not definitely a viable one.

	On the other hand, the claim that the probability distribution
should be cut off at the Planck density is at least equally {\it ad hoc}
and unjustified at present, so, just as one cannot yet be sure
that the total-probability solution does work,
neither can one be sure that it does not work.
In other words, it is not definitely not viable.

	A further argument against the total-probability solution
for saving the no-boundary proposal is the claim
that one-loop effects do indeed damp the probability density
in a calculable way \cite{Bar}, so one does not simply have to invoke
an {\it ad hoc} cutoff at the Planck density.
This may indeed be so for inflaton potentials below the Planck density,
but at the higher values of the inflaton field at nucleation,
where one needs to know the probability density,
one would expect the one-loop effects
to be dominated by an infinite sequence of yet higher-loop effects
that we do not know how to calculate.
(They are nonrenormalizable in ordinary Einsteinian gravity.)
Thus the one-loop calculations
can give only a hint of what may occur and do not tell the whole story
necessary to determine whether or not the total-probability argument
saves the no-boundary proposal.

	One case in which the total-probability argument definitely
seems not to work is the case in which the inflaton field has only a
(not too large) finite range \cite{PasPag}
(e.g., if its range is topologically $S^1$ rather than $R^1$)
or has only a finite volume
(e.g., as is the case for certain moduli spaces \cite{HM}).
If this could be proved to be the case
for the actual inflaton driving inflation,
then either one would need to look for another correction factor
or else conclude that the no-boundary proposal is inconsistent
with observations (or at least with our observations' being typical
observations predicted by the theory,
which is usually the best one can do with a quantum theory).

	A third correction is the use of the selection principle
called the weak anthropic principle (perhaps somewhat misleadingly,
since it is not meant to refer just to mankind) \cite{anth},
that what we observe about the universe is conditioned
on where we as observers exist within the universe.
Here I will take the ``where" to mean not only
where we are spatially within the universe
(e.g., on a planet with density at least 29 orders of magnitude
larger than the average density in a large neighborhood containing us),
and not only where we are temporally within the history of the universe
(e.g., at roughly $10^{61}$ Planck times after the beginning,
which is very long in Planck times but which may be very short
compared with the total lifetime of our region of the universe
if a superabundant amount of inflation occurred),
but also where we are within the quantum state of the universe
(e.g., where we are within the probability distribution
for different universe configurations).

	One obvious use of the weak anthropic principle is
the observation that the minimal amount of inflation is not
that which can barely be given the name inflation
(or lead to a quasi-classical universe \cite{GR}),
but is rather that which can lead to observers like us.
If observers require stellar nucleosynthesis
and/or heat from a nearby star,
which occurs for timescales around $10^{61}$ Planck times
(at least for the low energy dynamics of our component of the universe),
then as observers we would expect to be in a universe configuration
in which this long history occurred, which would give
a larger minimal value of inflation.

	However, this use of the weak anthropic principle
is not sufficient by itself to save the no-boundary proposal
from refutation by observation.
For even after restricting attention to the locations
within the quantum state of the universe where observers can occur,
we can still ask whether what we observe is typical of what
the distribution of observers within the theory do observe.

	For example, suppose that for the sake of argument
we assume what cannot yet be confirmed or refuted by calculation,
namely that the no-boundary probability density
for the nucleating value of the inflaton potential $V(\phi_0)$
is indeed damped or cut off above the Planck density,
so that the integral of the probability density
in the homogeneous minisuperspace approximation
is finite and is almost entirely dominated
by the exponentially large peak
at values of the potential that give only small amounts of inflation.
Now the weak anthropic principle would imply
that nearly all of this peak occurs for universes
that do not last long enough to have observers,
so what is relevant for observers
is the part of the probability distribution
above some observer-required minimum
of the nucleating inflaton value
$\phi_0$ that is significantly
to the right of the main part of the peak.
Here the probability density is far lower
than it is at the top of the peak,
but it is still falling very rapidly
with increasing nucleating inflaton field
(at least for typical forms of the inflaton potential,
such as a power law with a small coupling constant).

	Thus if we cut off the probability distribution
at values of the nucleating inflaton field $\phi_0$
below the observer-required minimum,
as well as above the field value that gives the Planck density,
the bulk of the probability integrated over values
between these two cutoffs
would occur for values just slightly above the lower cutoff.
These values of the inflaton field would lead to inflation
just slightly more than sufficient to give a universe
that lasts long enough for observers.
The bare minimum amount of inflation would have
the observers' existence occurring
near the end of a recollapsing universe,
just before it got too hot for them
(assuming that they mainly required
a minimum age of the universe for
star formation and/or nucleosynthesis).
Then the typical values of the the nucleating inflaton field
would lead to observers' existing just slightly earlier
in a recollapsing universe, clearly in conflict with
our observation that the universe is still expanding.

	However, the weak anthropic principle
(or any similar selection principle that postulates
that we are typical conscious beings, observers,
or parts of a civilization)
is capable of saving the no-boundary proposal
when one considers inhomogeneous
inflaton fluctuations and their back-reaction on the metric,
a process called `stochastic inflation,' which
can lead to a `self-reproducing universe'
with `eternal inflation'
that occurs for an indefinitely long time
and hence makes the volume of the universe
arbitrarily large
\cite{V2,St,EL,AV,Mi,LLM}.
The idea is based on the observation
that typical observers or civilizations are more likely
to occur in spaces of larger volume,
other factors being equal \cite{EL,AV,Lin90,LLM,V95}.

	The basic idea of eternal inflation was
present in the original picture of `old' inflation
by Guth \cite{Guth}, in which inflation occurs forever
in the `false vacuum' region that grows indefinitely.
In that model eternal inflation was a problem,
because the bubbles of `true vacuum' that form
were essentially empty inside (not suitable
precursors of our part of the universe),
and collisions between bubble walls were
too infrequent to percolate and lead to
a thermalized and fairly homogeneous region
\cite{HMS,GW}.  This `graceful exit' problem 
was cured by `new' inflation \cite{LNI,AS}
(inflation starting at or near
a maximum of an inflaton potential),
in which the inflaton potential is sufficiently
flat that the inflaton undergoes a `slow roll'
down it, leading to
a thermalized and fairly homogeneous region
within each bubble at the end of inflation. 
Steinhardt \cite{Stein} noted that 
new inflation can lead to a
`regenerative meta-cosmology'
with new bubbles regenerated forever
by inflation outside the bubbles
(as in Guth's old inflation, except that in new inflation
the bubbles are not empty and so are possible
precursors for our present region of the universe).
Linde \cite{LCam} followed up this idea by suggesting
that the universe might be in this stationary
`self-reproducing' state forever in the past
as well as in the future, 
thus eliminating the need for an initial singularity.

	The first application of the inhomogeneous
stochastic evolution of the inflaton field to
eternal inflation was made by Vilenkin \cite{V2} for
new inflation, when he showed that one could get
an arbitrarily large amount of inflation even within a bubble 
(which could encompass the entire universe in Vilenkin's picture, 
as in that of Hawking and Moss \cite{HawMos}), 
and not just outside the bubbles as in previous analyses.
After Linde discovered the
scenario of `chaotic inflation' \cite{LCI}
(inflation from an inflaton potential without
a maximum, or in a region where there is no maximum),
which seems more realistic than new inflation,
he discovered that it also leads to
eternal stochastic inflation
and what he called a `self-regenerating universe'
\cite{EL,Lin90}.  (Linde was actually the first to use
the phrase `eternal inflation' \cite{EL}, and he has
been the leading researcher of it since that time
\cite{EL,Lin90,LLM}.) 

	Eternal stochastic inflation occurs when the rms
`stochastic' change in the scalar field from the
freezing out of inhomogeneous modes during one
Hubble time $\Delta t = H^{-1}$,
 \begin{equation}
 \delta\phi = {H \over 2\pi} = \sqrt{2V \over 3\pi},
 \label{eq:27}
 \end{equation}
is greater than the slow-roll change of the
field during that same time,
 \begin{equation}
 \Delta\phi = -\dot{\phi}\Delta t
 = {V' \over 3H^2} = {V' \over 8\pi V}.
 \label{eq:28}
 \end{equation}
This occurs when the stochastic inflation condition
 \begin{equation}
 V'^2 < {128 \pi \over 3} V^3,
 \label{eq:29}
 \end{equation}
is satisfied \cite{Lin90}.

	As a result of both the stochastic and slow-roll
changes in the inflaton field, in some regions
the field decreases, and in others it increases.
Although the amount of comoving volume in which
the field decreases is greater than the amount in which
the field increases (because of the slow-roll change
$\Delta\phi$, which is toward smaller fields),
the back-reaction of the inflaton potential on the metric
causes the physical volume to increase more in the regions
in which the field increases.  Therefore, when one weights
the regions by their physical volumes rather than by
their comoving volumes, the dominant behavior
is for the inflaton field to increase.  This process allows
the inflaton field to remain large for an arbitrarily long time,
thereby leading to an arbitrarily large amount of inflation
\cite{V2,EL,AV,Mi,LLM}.
The results of eternal stochastic inflation
are claimed to be independent of the initial conditions
\cite{LLM}
(a claim which seems to me implicitly to assume
some strong restriction on the allowed quantum states
but which apparently allows both the no-boundary
and tunneling states).

	Now the main point of the present paper
is the conceptual or pedagogical point that
even at the crude level
of using only the zero-loop homogeneous
minisuperspace approximation,
the anthropic-principle idea of weighting
by the physical volume can save the Hartle-Hawking
no-boundary proposal from appearing inconsistent with
our observations of an expanding universe,
at least for a wide range of inflaton potentials,
even if the probability distribution
is damped or cut off at the Planck density.

	To use Vilenkin's language \cite{V95},
suppose we start
with what he calls the ``principle of mediocrity,"
that our civilization is average,
``randomly picked in the metauniverse."
This leads to a probability distribution
for various observed results
that is proportional not only to what I shall call
the `bare' probability distribution of universe configurations
having these results, but also to the number of civilizations
occurring within the corresponding configuration.
(Note that what I am calling a universe configuration,
Vilenkin calls a universe, and what I call the universe,
Vilenkin calls the metauniverse.)

	The bare probability distribution is
that given by the appropriate probability interpretation
for the correct quantum state of the universe
that does not make reference to conscious beings,
observers, or civilizations.
(This interpretation, or even its existence, is not yet agreed upon,
but perhaps it does exist and give a probability density
something like the absolute square of the wavefunction,
or maybe proportional
to an appropriate Klein-Gordon flux in superspace).
Then, with other factors being equal, one would expect
the number of civilizations to be proportional to
the volume of space at the time at which the civilizations occur.
(I myself might prefer \cite{SQM} to focus on conscious perceptions
rather than civilizations, and someone else might prefer
to focus on observers, but one would expect any of these
to be proportional to the volume of space,
other conditions being the same.)
The volume of space at the time of the civilizations would itself
be proportional to the volume of space at the end of inflation
(assuming that inflation occurred,
and that there is a fixed volume expansion factor
between the end of inflation and the time of the civilizations,
as there would be for approximately k=0 Friedmann-Robertson-Walker
parts of the universe with the same density at the end of inflation,
the same density at the time of the civilizations,
or the same post-inflation age then,
and the same equation of state at the intermediate densities).

	Thus one would expect that the probability distribution
for observed results to be roughly proportional
to the bare probability distribution for these results,
multiplied by the volume of space at the end of inflation
in the universe configuration that has these results.
I shall call this product
the `observational' probability distribution.

	(In some quantum theories,
such as my own Sensible Quantum Mechanics \cite{SQM},
the observational probability distribution may be
the only probability distribution given by the theory,
and the bare probability distribution may be undefined,
but if this is so, I can still employ in this paper
the useful fiction that the observational probability distribution
factorizes into a product of a bare probability distribution
for some physical configuration or history,
and of a probability or measure of how much
of certain types of observations occur
within that configuration or history.
Also, as in Sensible Quantum Mechanics
and in other variations
of the many-worlds formulation \cite{MW},
the probability distribution may actually be a measure distribution
on an actual ensemble predicted by the theory,
rather than any description of a random choice,
not precisely determined by the theory,
of a single actual result out of a hypothetical ensemble
of possible results, as in versions of quantum mechanics
in which the quantum state is occasionally supposed to collapse
in some particular way not uniquely specified by the theory.)

	Here I shall focus on the probability distribution,
in the minisuperspace approximation,
of one `constant of motion,' $\phi_0$,
of an approximate classical universe model
that matches a universe configuration.
(The probability distribution of $\phi_0$
in stochastic inflation has been studied in \cite{LLM}.)
In particular, I shall focus on universe configurations
in which the effective constants of nature are the same as
in our configuration and in which the large-scale configuration
is approximately that of a classical 
$k=1$ Friedmann-Robertson-Walker
universe which evolved from a period of single inflation
starting with a moment of time symmetry
at which the size of the universe was a minimum
and the inflaton field had the homogeneous value $\phi_0$.

	According to the Hartle-Hawking no-boundary proposal
in the minisuperspace approximation being used here,
the zero-loop approximation for the bare probability distribution
gives the unnormalized approximate probability density
of Eq. (\ref{eq:1}) for $\phi_0$,
but with $P_{\rm NB}$ there replaced by $P_{\rm bare}$
here to emphasize that it is the approximate bare probability,
namely,
 \begin{equation}
 P_{\rm bare}(\phi_0) d\phi_0 = e^{\pi a_0^2(\phi_0)} d\phi_0
	= \exp{[{3 \over 8V(\phi_0)}]} d\phi_0.
 \label{eq:4}
 \end{equation}
(By ``approximate," I mean, e.g., ignoring multi-loop effects,
Jacobians, or other prefactors of the exponential
of twice the negative of the real part of the Euclidean action.
I also mean that this Euclidean action is itself given here
only in the approximation that variations in the potential
and in the energy density are negligible
during a Euclidean regime that is assumed to be a
Friedmann-Robertson-Walker 4-dimensional hemisphere
of radius $a_0$ given by Eq. (\ref{eq:3}),
bounded by a totally-geodesic round equatorial 3-sphere 
that is the 3-space of the universe
at its moment of nucleation out of the Euclidean regime
and into the Lorentzian regime of inflation.)

	As noted above, the bare probability
rises sharply for smaller values of $V(\phi_0)$
when this is much smaller than unity (the Planck density).
However, we need to multiply the bare probability
by the volume of space at the end of inflation
to get the observational probability.

	The volume of 3-space at the moment of nucleation,
which is the beginning of a Lorentzian period of inflation
that would have had a moment of time symmetry then
if the Lorentzian evolution were analytically continued 
backward as well as forward
in real Lorentzian time from this moment of nucleation, is
 \begin{equation}
 {\cal V}_0 = 2\pi^2 a_0^3 = [{27\pi \over 128 V^3(\phi_0)}]^{1/2}.
 \label{eq:5}
 \end{equation}

	Then we need to multiply
by the volume expansion factor during inflation.
Let us assume that the inflaton potential has
$0 < V'(\phi) \equiv dV/d\phi << V(\phi) <<1$
for $\phi_{\rm e} << \phi << \phi_{\rm Pl}$,
with $\phi_{\rm e}$ being the value of $\phi$ where
 \begin{equation}
 V'(\phi_{\rm e}) = V(\phi_{\rm e})
 \label{eq:5b}
 \end{equation}
(which will be taken to be the point at which inflation ends,
since this is roughly the point 
at which the slow-roll approximation breaks down),
and with $\phi_{\rm Pl}$ being the value
of the inflaton field that leads to a potential of the Planck density,
 \begin{equation}
 V(\phi_{\rm Pl}) = 1
 \label{eq:5c}
 \end{equation}
in the Planck units I am using.
Such a potential leads to slow-roll inflation for $\phi$ in this range
(which is assumed to include $\phi_0$).
For simplicity, use the slow-roll approximation all the way to the end
of inflation at $\phi = \phi_{\rm e}$.
Then the slow-roll approximations to the FRW-inflaton equations
lead to a volume expansion factor during inflation of roughly
 \begin{equation}
 {{\cal V}_{\rm e} \over {\cal V}_0} = ({a_{\rm e} \over a_0})^3
 = \exp{[24\pi\int_{\phi_{\rm e}}^{\phi_0}{V(\phi)d\phi \over V'(\phi)}]}
 = \exp{[12\pi\int_{\phi_0}^{\phi_{\rm e}}{a(\phi)d\phi \over a'(\phi)}]}.
 \label{eq:6}
 \end{equation}

	Multiplying this volume expansion factor
by the volume of space at the beginning of inflation
and by the bare probability distribution gives
the unnormalized observational probability distribution
 \begin{eqnarray}
 P_{\rm obs}(\phi_0) d\phi_0
	&=& {\cal V}_{\rm e} P_{\rm bare}(\phi_0) d\phi_0 \nonumber \\
	&=& {\cal V}_0 (\frac{a_{\rm e}}{a_0})^3 P_{\rm bare}(\phi_0) d\phi_0
	 \nonumber \\
	&=& [\frac{27\pi}{128 V^3(\phi_0)}]^{1/2}
	\exp{[24\pi\int_{\phi_{\rm e}}^{\phi_0}\frac{V(\phi)d\phi}{V'(\phi)}]}
	\exp{[\frac{3}{8V(\phi_0)}]} d\phi_0\nonumber \\
	&=& 2\pi^2 a_0^3(\phi_0)
 	\exp{[12\pi\int_{\phi_0}^{\phi_{\rm e}}\frac{a(\phi)d\phi }{ a'(\phi)}]}
	\exp{[\pi a_0^2(\phi_0)]}
	= e^{p(\phi_0)} d\phi_0,
 \label{eq:7}
 \end{eqnarray}
where the logarithm of
the unnormalized observational probability density is
 \begin{eqnarray}
 p(\phi_0)
	&=&  24\pi\int_{\phi_{\rm e}}^{\phi_0}{V(\phi_0) d\phi \over V'(\phi_0)}
	+ {3 \over  8V(\phi_0)}
	- {3 \over  2}\ln{V(\phi_0)} + {1 \over  2}\ln{({27\pi \over128})}
	\nonumber \\	
	&=& 12\pi\int_{\phi_0}^{\phi_{\rm e}}
		{a_0(\phi)d\phi \over a'_0(\phi)}
	+ \pi a_0^2(\phi_0) + 3\ln{a_0(\phi_0)} + \ln{(2\pi^2)}
	\nonumber \\	
	&\approx &
	24\pi\int_{\phi_{\rm e}}^{\phi_0}{V(\phi_0) d\phi \over V'(\phi_0)}
	+ {3 \over 8V(\phi_0)}
	= 12\pi\int_{\phi_0}^{\phi_{\rm e}}
		{a_0(\phi)d\phi \over a'_0(\phi)}
	+ \pi a_0^2(\phi_0).
 \label{eq:8}
 \end{eqnarray}
I shall generally use one of these last two approximate expressions,
dropping the logarithm of the volume at the beginning of inflation
as a relatively unimportant term,
and keeping only the logarithm of the volume expansion factor
during inflation and minus twice the Euclidean action
in the zero-loop approximation to the bare probability,
since those two terms generally dominate when $V(\phi_0) << 1$,
or, equivalently, when $a_0(\phi_0) >> 1$
(nucleating universe much larger than the Planck size).

	Now we need to put in the fact
that the observational probability density is cut off
for $\phi_0 < \phi_{\rm m}$, where $\phi_{\rm m}$ is the minimum value
of the nucleating inflaton field to lead to enough inflation
for the existence of civilizations.
If civilizations can only occur when the universe
is old enough for some nucleosynthesizing stars to have burned out
and yet for other heat-producing stars to still be burning,
then if the constants of nature take the values that they do
in our universe configuration,
one would need at least $N_{\rm m}$ (roughly 60 \cite{Lin90,KT})
$e$-folds of inflation,
and this makes $\phi_{\rm m}$ the solution of the following equation:
 \begin{equation}
 N_{\rm m} = 8\pi\int_{\phi_{\rm e}}^{\phi_{\rm m}}{V(\phi)d\phi \over V'(\phi)}.
 \label{eq:9}
 \end{equation}

	If, for the sake of argument,
we also cut off the probability distribution
for $\phi_0 > \phi_{\rm M}$ (e.g., with $\phi_{\rm M} = \phi_{\rm Pl}$),
then unless $\phi_{\rm M}$ is utterly enormous
(which would require that $V(\phi)$ be extremely flat
at large $\phi$ if $V(\phi_{\rm M}) \leq 1$),
then most of the total (integrated) observational probability
for $\phi_{\rm m} < \phi_0 < \phi_{\rm M}$
will occur near the maximum
of the approximate unnormalized observational probability density,
or of its logarithm $p(\phi_0)$.
If the maximum is at or very near $\phi_{\rm m}$,
then one would predict that a typical civilization
would see the universe recollapsing,
which is contrary to our observations.
But if the maximum is at a sufficiently higher value of $\phi_0$,
there would be enough inflation that the universe today
would be much larger than what we can see
and hence very nearly spatially flat
on a scale corresponding to its present age
(under the assumption that it is
approximately Friedmann-Robertson-Walker),
thus agreeing with observations.

	Now the analysis depends on the qualitative form
of the inflaton potential $V(\phi)$.
Because a period of slow-roll inflation
has $V(\phi)$ monotonically decreasing with time
while $\phi$ itself also changes monotonically,
I shall assume that within the entire range
$\phi_{\rm m} < \phi < \phi_{\rm M}$,
$V'(\phi)$ is bounded away from zero and hence has a single sign
(which without loss of generality is herein taken to be positive,
since one could replace $\phi$ by $-\phi$ if necessary).
I shall also assume that $V(\phi)$ is sufficiently smooth
to be at least twice differentiable as a function of $\phi$.

	The first two derivatives of the approximate expression for the
logarithm of the probability density, $p(\phi_0)$, then have the form
 \begin{equation}
 p' \equiv {dp \over d\phi_0} = {24\pi V \over V'} - {3V' \over 8V^2}
	= 2\pi a_0 (a'_0 - {6 \over a'_0}),
 \label{eq:10}
 \end{equation}
 \begin{eqnarray}
 p'' \equiv {d^2 p \over d\phi_0^2}
 &=& 24\pi (1 - {V V''\over V'^2})
 - {3 \over 8}({V'' \over V^2} - 2{V'^2 \over V^3})\nonumber \\
 &=& 2\pi [a_0 (1 + {6 \over a'^2_0})a''_0 + a'^2_0 - 6]
	\nonumber \\
 &=& 2\pi a_0 (1 + {6 \over a'^2_0})a''_0 + {a'_0 \over a_0} p',
 \label{eq:11}
 \end{eqnarray}
where the primes on the $V$'s and $a_0$'s, just as on the $p$'s,
mean derivatives with respect to the $\phi_0$ independent variable,
of which they are functions.

	For a fairly general class of potentials $V(\phi_0)$,
which I shall call Class 1, $p(\phi_0)$ has no local maximum between
$\phi_{\rm m}$ and $\phi_{\rm M}$.  For example, at a local extremum of $p$,
one can see from Eqs. (\ref{eq:10}) and (\ref{eq:11}) that
the second derivative of $p$ with respect to $\phi_0$
has the same sign as the second derivative of $a_0$ or of $V^{-1/2}$,
so if these functions are concave upward
(as they are, for example, if $V$ is a positive power of $\phi_0$
or is exponentially increasing with $\phi_0$),
then $p(\phi_0)$ has no local maximum.

	Within this Class 1, the maximum of $p(\phi_0)$
occurs at one of the endpoints, at $\phi_{\rm m}$ or $\phi_{\rm M}$,
and so it is simply a question of whether 
$p(\phi_{\rm m})$ or $p(\phi_{\rm M})$ is larger, 
assuming that the difference is greater
than the generally less-important factors we have dropped.
If $p(\phi_{\rm m})$ is the larger,
then the observational probability for the Hartle-Hawking proposal
(at least within the minisuperspace and zero-loop approximations)
would be dominated by cases in which observers occurred
almost entirely very late within a recollapsing universe,
which is contrary to observations.
But if $p(\phi_{\rm M})$ is the larger,
then the observational probability would be dominated
by cases in which the universe is expanding
near the $k=0$ borderline when observers occur within it,
which is consistent with our observations.

	For the complementary class, which I shall call Class 2,
$p(\phi_0)$ does have one or more local maxima
between $\phi_{\rm m}$ and $\phi_{\rm M}$,
where $a'_0 = -\sqrt{6}$ and $a''_0 < 0$.
In this case one needs to compare the values of $p(\phi_0)$
at these local maxima as well as at the endpoints,
$\phi_{\rm m}$ and $\phi_{\rm M}$.
Still assuming that ignored factors are insignificant,
and assuming that no local maximum occurs
so close to $\phi_{\rm m}$ that it would give insufficient inflation
to be consistent with observations,
the only case in which the Hartle-Hawking proposal
would apparently (i.e., if our approximations are valid)
give typical results inconsistent
with our observations of the expansion of the universe
would be the case in which $p(\phi_{\rm m})$ is larger than the value of
$p(\phi_0)$ at the other endpoint
or any of the local maxima in between.

	One {\it sufficient} (but not {\it necessary}) condition
for $p(\phi_{\rm m})$ not to be the global maximum of $p(\phi_0)$
for $\phi_{\rm m} < \phi_0 < \phi_{\rm M}$,
and hence for the Hartle-Hawking proposal to be consistent
with our observations of the expansion of the universe
(assuming, as always, that other corrections factors are negligible),
is that $p'(\phi_{\rm m}) > 0$.
In terms of the potential $V$
and its derivative $V'$,
the sufficient condition for the Hartle-Hawking proposal
to pass this test is
 \begin{equation}
 V'^2 < 64 \pi V^3,
 \label{eq:12}
 \end{equation}
when evaluated at $\phi_0 = \phi_{\rm m}$.
Or, in terms of the derivative $a'_0$ (still with respect to $\phi_0$)
of $a_0 = (8\pi V/3)^{-1/2}$
at $\phi_0 = \phi_{\rm m}$, it is
 \begin{equation}
 -\sqrt{6} < a'_0 < 0.
 \label{eq:13}
 \end{equation}

	On the other hand, for $p(\phi_0)$
to be greater than $p(\phi_{\rm m})$
for some larger value of $\phi_0$,
one has the {\it necessary} condition
that Eq. (\ref{eq:12}) or (\ref{eq:13})
be true when evaluated in at least some range of
$\phi_0$ greater than $\phi_{\rm m}$.

	Consider the Class 1 example of a power-law potential
with positive (constant) exponent $n$,
 \begin{equation}
 V(\phi) = {\lambda \over n} \phi^n,
 \label{eq:14}
 \end{equation}
where $\lambda$ is a coupling constant for the field,
which in the Planck units we are using is
a number that shall be assumed to be small.
(For example, for $n = 2$, it is the square of the mass
of the inflaton field $\phi$,
which is then a free massive field,
minimally coupled to gravity.)

	Again making the approximation of keeping only
the volume expansion factor and
the zero-loop bare probability factor,
this leads to the logarithm
of the observational probability density varying roughly as
 \begin{equation}
 p(\phi_0) = p(\phi_{\rm m}) + {12\pi \over n}(\phi_0^2 - \phi_{\rm m}^2)
 - {3n \over 8\lambda}(\phi_{\rm m}^{-n} - \phi_0^{-n}).
 \label{eq:15}
 \end{equation}
For $\phi_0 >> \phi_{\rm m}$, one gets, roughly,
 \begin{equation}
 p(\phi_0) - p(\phi_{\rm m})
 = {12\pi \over n}\phi_0^2  - {3n \over 8\lambda \phi_{\rm m}^n}.
 \label{eq:16}
 \end{equation}
Since there are are no local maxima of $p(\phi_0)$ for this Class 1
potential, it would allow the Hartle-Hawking proposal
(in the minisuperspace approximation under consideration)
to be consistent with observations if and only if
$p(\phi_{\rm M}) > p(\phi_{\rm m})$, or, roughly,
 \begin{equation}
 \phi_{\rm M}^2 > {n^2 \over 32\pi\lambda \phi_{\rm m}^n}.
 \label{eq:17}
 \end{equation}

	To express this condition as a condition
on the coupling constant $\lambda$ for a given exponent $n$,
we need to write $\phi_{\rm M}$ and $\phi_{\rm m}$
in terms of these two parameters of the potential.  If one takes
$\phi_{\rm M} = \phi_{\rm Pl}$, the value where $V = 1$, one gets
 \begin{equation}
 \phi_{\rm M} = \phi_{\rm Pl} = ({n \over \lambda})^{1 \over n}.
 \label{eq:18}
 \end{equation}
Furthermore, $\phi_{\rm m}$ is determined by the need for $N_{\rm m}$
(roughly 60 \cite{Lin90,KT}) $e$-folds of inflation
before $\phi$ decreases to $\phi_{\rm e}$,
where the slow-roll approximation ends and inflation ends.
For the power-law potential given by Eq. (\ref{eq:14}),
Eq. (\ref{eq:5c}) gives
 \begin{equation}
 \phi_{\rm e} = n,
 \label{eq:19}
 \end{equation}
and then for $N_{\rm m}$ $e$-folds of inflation, we need
 \begin{equation}
 N_{\rm m} = \ln{(a_{\rm e}/a_0)} = {4\pi \over n}(\phi_{\rm m}^2 - \phi_{\rm e}^2),
 \label{eq:20}
 \end{equation}so
 \begin{equation}
 \phi_{\rm m} = \sqrt{\phi_{\rm e}^2 + {n N_{\rm m} \over 4\pi}}
 = \sqrt{n^2 + {n N_{\rm m} \over 4\pi}}
 \approx ({n N_{\rm m} \over 4\pi})^{1/2} \sim 2\sqrt{n},
 \label{eq:21}
 \end{equation}where the first approximation is for $n << N_{\rm m}/4\pi$,
and the second \cite{Lin90} uses the fact that $N_{\rm m}/4\pi \sim 4$.
(The fact that this number is not very large suggests that
even the first approximation is not very good, but in a more careful
analysis \cite{Lin90} the slow-roll approximation breaks down at a
$\phi_{\rm e}$ that is actually something like $n/\sqrt{16\pi}$,
which would make the first term inside the square root of
Eq.(\ref{eq:21}) about 50 times smaller than the crude estimate above.)

	Now if one inserts $\phi_{\rm M}$ from Eq. (\ref{eq:18}) and
the last approximation for $\phi_{\rm m}$ from Eq. (\ref{eq:21})
into the inequality (\ref{eq:17}), one finds that it becomes
 \begin{equation}
 \lambda^{n-2 \over n} > 2^{-n(n+5) \over 2} \pi^{-{n \over 2}}
	n^{-({n-2 \over 2})^2},
 \label{eq:22}
 \end{equation}
the condition for the Hartle-Hawking proposal
(in the minisuperspace and zero-loop approximations)
to have most observers see a nearly flat universe,
consistent with our observations, rather than
a recollapsing universe,
for the power-law potential (\ref{eq:14}).

	We have already assumed that $\phi_{\rm m} << \phi_{\rm M}$,
which implies that the coupling constant must be small,
 \begin{equation}
 \lambda << 2^{-n} n^{2-n \over n},
 \label{eq:23}
 \end{equation}
so the inequality (\ref{eq:22}) is automatically true for $n \leq 2$
(recall that we are assuming that $n > 0$).
In particular, for a free massive inflaton ($n=2$),
with mass much less than the Planck mass,
the Hartle-Hawking proposal
(even at the minisuperspace level being considered)
would be consistent
with our observations in predicting that a typical observer
would see a nearly flat universe on large scales.

	However, for exponents $n > 2$, the consistency
of the Hartle-Hawking proposal is not automatic
at this minisuperspace level.
There is always a range of values of the coupling constant
$\lambda$ that is consistent with both inequalities
(\ref{eq:22}) and (\ref{eq:23}), but for sufficiently
large values of $n$, the allowed range for $\lambda$
is at values too large to be consistent with the observed density
fluctuations of the universe (which one can calculate
only by going outside the minisuperspace approximation,
at least for the fluctuations).

	For example, one may use the approximate expression
Linde \cite{Lin90} gives (on p. 185) for the coupling constant of a
power-law potential from the density fluctuations of the universe,
 \begin{equation}
 \lambda \sim 2.5\cdot 10^{-13}n^2(4n)^{-n/2}.
 \label{eq:24}
 \end{equation}
Then one can readily calculate, using the approximations above,
 \begin{equation}
 p(\phi_0) - p(\phi_{\rm m})
 \approx {12\pi \over n}\phi_{\rm M}^2  - {3n \over 8\lambda \phi_{\rm m}^n}
 \sim 48\pi({4\cdot 10^{12} \over n})^{2 \over n}
	- {3 \over 8}({4\cdot 10^{12} \over n}),
 \label{eq:25}
 \end{equation}
and this last expression is positive if and only if
 \begin{equation}
 n < 2.5430075348.
 \label{eq:26}
 \end{equation}
Of course, the crudeness of the approximations above
do not justify the precision given here for the value of $n$
at which the last expression of Eq. (\ref{eq:25}) vanishes;
it merely suggests that for $n$ greater than roughly $5/2$,
the minisuperspace and zero-loop approximations seem to make
the Hartle-Hawking proposal be in conflict with
observations if one cuts off the distribution
of nucleating universes at the Planck density.
Such a conflict does not occur for any power-law
potential with a suitably small coupling constant $\lambda$
if one goes beyond the minisuperspace approximation
to eternal stochastic inflation \cite{LLM}.

	Thus we see that for a power-law potential,
when one includes the volume factor in the
distribution of observers
(or of civilizations, or simply of conscious beings),
the minisuperspace and zero-loop approximations
for the Hartle-Hawking no-boundary proposal
give results consistent with our observations
of a universe expanding near the critical density,
even when an {\it ad hoc} cutoff is imposed
on the minisuperspace toy model at the Planck density,
if the exponent of the power-law is smaller than roughly
2.5.  This includes the simple case of a free massive
field but excludes the case of a quartic potential
(though the latter is allowed in inhomogeneous
models giving eternal stochastic inflation \cite{LLM}).

	Of course, there are other forms of the potential
that would also make the Hartle-Hawking proposal
consistent with observations by the approximations above.
These include the cases in which the potential has a
smooth maximum (below the Planck density)
at some finite field value,
and the case in which the potential continues
to rise for arbitrarily large field values but
asymptotically approaches a finite limit
(also assumed to be below the Plank density
so that no cutoff need be made).

	In both of these cases, one can get an
arbitrarily large volume by having the field
nucleate arbitrarily near the maximum of the potential
in the first case, or at an arbitrarily large field
value in the second case.  Then the slow-roll approximation
will give an arbitrarily large amount of inflation,
so the volume factor can become arbitrarily large
and hence dominate over any large (but necessarily finite)
peak in the bare probability distribution.
(For this peak to be finite, I am assuming that
the cosmological constant is zero or positive, so that
the potential is bounded below by zero, and that
inflation occurs only when the potential has
a positive value, strictly bounded away from zero.)

	It is interesting that the inequality (\ref{eq:12}) is the same
(up to a small change in the coefficient that is not important
at the level of the approximations being employed here)
as the inequality (\ref{eq:29}) that occurs for some range within
an inflaton potential allowing eternal inflation.
Therefore, when the Hartle-Hawking approximation
in the minisuperspace approximation
is consistent with our observations of an expanding universe,
then at the level of considering inhomogeneous fluctuations,
it leads to stochastic inflation and a large expanding
universe also consistent with such observations.
However, the converse is not true, since potentials
obeying the inequality (\ref{eq:29}) somewhere within
the allowed range, and leading to stochastic inflation
within this range, need not in the minisuperspace
approximation necessarily have the peak
in the observational probability distribution be at a nucleating
inflaton value $\phi_0$ that is higher than the minimum
value $\phi_{\rm m}$.  (The inequality (\ref{eq:29}) merely
implies that the observational probability density is
rising with $\phi_0$ there, but not that it is necessarily
higher there than it is at $\phi_{\rm m}$.)

	In particular, for any power-law potential (\ref{eq:14})
with a small coupling constant $\lambda$,
the inequality (\ref{eq:29}) is true for \cite{Lin90}
 \begin{equation}
 \phi > ({3n^3 \over 128\pi\lambda})^{1 \over n+2}
 \label{eq:30}
 \end{equation}
(which is a value below $\phi_{\rm Pl}$ and hence within
the allowed range if $\phi_{\rm M} \geq \phi_{\rm Pl}$).
Thus stochastic inflation can occur for any power-law
potential (with a positive exponent and a sufficiently
small coupling constant), giving a Hartle-Hawking state
consistent with our observations of a large expanding universe,
even though the misuperspace approximation,
for exponents $n$ greater than about 2.5,
would suggest that the state would give typical observations
of a recollapsing universe, if one cut off the probability
distribution at the Planck density $V(\phi_{\rm Pl}) = 1$.

	In summary, when one includes the volume of space
in converting from bare probabilities to observational
probabilities, then the Hartle-Hawking no-boundary proposal
for the quantum state of the universe, as well as the
tunneling proposal, both seem to have enough space to be
consistent with our observations of a nearly flat expanding
universe (rather than a contracting universe), at least for
a wide class of inflaton potentials
that obey the inequality (\ref{eq:29}) somewhere
within the allowed range of the inflaton field,
even if one cuts off the probability distribution
for universes nucleating above the Planck density.
This fact has been known to be the case for
eternal stochastic inflation \cite{Mi,LLM}, and here
the pedagogical point is made
that the consistency of both proposals with
the aforementioned observations occurs
even within the minisuperspace approximation
for a certain subset of the potentials
that allow eternal inflation
(e.g., for a massive scalar field,
though not for a quartic potential,
despite the fact that the latter does allow
eternal stochastic inflation,
and hence consistency with observations in the realistic case
in which one allows inhomogeneous metrics).

	Therefore, when one tests theories of quantum cosmology
against our observations of a large expanding universe,
there is often space for both the no-boundary
and the tunneling proposals,
even within the zero-loop minisuperspace approximation.

	On the other hand, there are inflaton potentials
(such as the power-law potentials with exponents
larger than roughly 2.5) that would
make the Hartle-Hawking no-boundary proposal,
at the zero-loop minisuperspace level
with the probability distribution
for nucleating universes
cut off at the Planck density,
appear to be inconsistent with our observations
of an expanding universe,
even though a calculation invoking
eternal stochastic inflation
would show that it is actually consistent.

	One might ask whether it is the zero-loop
or the minisuperspace approximation (or both)
that in these cases makes such a large difference
from eternal stochastic inflation.
I would conjecture that although stochastic
inflation requires one to go beyond the homogeneous
minisuperspace approximation, it may not require one
to go beyond the zero-loop approximation.
Very preliminary evidence suggests to me that
one should be able to get something like stochastic
inflation simply be considering inhomogeneous complex
classical solutions of the Einstein-matter field equations
that obey the no-boundary conditions (when these conditions
are expressed as analytic equations that may be satisfied
by complex solutions).  In the zero-loop approximation,
the bare probabilities would then be given simply
by the exponential of minus twice the real part of
the Euclidean action (the imaginary part of the Lorentzian action),
but then to get the observational probabilities
one would need to multiply by the volume
of space on hypersurfaces where the local conditions
are suitable for civilizations, observers, or conscious beings.
Further details of this will be left for another paper \cite{P97}.

	I am less certain how to apply the tunneling
conditions to inhomogeneous complex
classical solutions to get a zero-loop approximation
for that proposal, but if one can do it, I would
expect that the results for large universes would
be virtually the same as for the no-boundary proposal.
As stated in \cite{LLM}, "the indefinitely large growth
of volume produced by inflation does solve
the problem of initial conditions for every theory
where the self-reproduction of the universe is possible."

	I should also point out that I do not believe that the
results would be indistinguishable for {\it all} quantum states
of the universe, but only those that lead to self-reproduction
with separate Hubble-sized regions being sufficiently uncorrelated.
Although this restriction would allow a large class of quantum
states that lead to eternal stochastic inflation and perhaps
locally indistinguishable observations in typical post-inflation
epochs where life may occur, it would presumably exclude an even
larger class of quantum states.

	An analogy in nongravitational quantum field theory
in classical Minkowski spacetime
would be the set of suitably nonpathological Fock states
in which the excitations disperse with time,
so that in each local spatial region,
the quantum state asymptotically approaches
that of the vacuum.
There is a large class of such QFT states,
but they are a restrictive set of measure zero
compared to a more general class of states
that never asymptotically approach the vacuum
in any local spatial regions
(e.g., perturbations of thermal states
at any nonzero temperature,
a real parameter that can have a continuum of values).

	Therefore, even if it turns out that
our observations cannot distinguish between
the no-boundary proposal, the tunneling proposal,
and perhaps any other among a certain large class of proposals,
the remarkable order that we observe
strongly suggests to me that the quantum state
of the universe is highly special.
It is tempting to speculate that it might
be extremely simple.
The no-boundary and tunneling proposals
are sketches for two very simple quantum states.
Although for the goal of discovering the precise
quantum state of the universe,
it is discouraging that both of these
proposals seem (at least so far)
to predict indistinguishable observations,
it is encouraging that (again, at least so far as we now know)
both seem to be consistent with our observations.

	Appreciation is expressed for the hospitality
of the Ettore Majorana Centre for Scientific Culture in
Erice, Sicily, where at the International School of Astrophysics
``D. Chalonge" on String Gravity and Physics at the Planck Scale
in September 1995, lectures by, and conversations with,
Alex Vilenkin led me to the main idea expressed here.
At a February 1997 CIAR conference in Banff, Alberta,
Andrei Linde gave me a very helpful summary
of the ideas of eternal stochastic inflation
and their history.
E-mail discussions with Linde, Vilenkin, and Bob Wald
have also been very useful.
This research was supported in part by
the Natural Sciences and Engineering Research Council of Canada.

\end{document}